\documentclass{article}
\usepackage{graphicx,subcaption,lipsum} 
\usepackage[hmargin=2.7cm,vmargin=3cm]{geometry}
\usepackage[utf8]{inputenc}
\usepackage{amsmath}
\usepackage{float}
\usepackage{enumerate}
\usepackage{amsfonts}
\usepackage{physics}
\usepackage{mathtools}
\usepackage{appendix}
\usepackage{multirow}
\newcommand{\R}{\mathbb{R}}

\newcommand{\Z}{\mathbb{Z}}
\newcommand{\vfi}{\varphi}
\DeclareMathOperator{\sinc}{sinc}

\usepackage{cite}
\usepackage[colorlinks,pagebackref,pdfusetitle,urlcolor=blue,citecolor=blue,linkcolor=blue,bookmarksnumbered,plainpages=false]{hyperref}

\usepackage{graphicx,color}
\usepackage{float}
\usepackage{epstopdf}
\newcommand{\bey}{\begin{eqnarray}}
\newcommand{\eey}{\end{eqnarray}}
\usepackage{microtype}
\usepackage{xcolor}

\begin{document}
\title{Time of arrival on a ring and relativistic quantum clocks}
\author {Iason Vakondios\texorpdfstring{\thanks{i.vakondios@ac.upatras.gr}}{}\texorpdfstring{\;}{ } and \texorpdfstring{\;}{ }Charis Anastopoulos\texorpdfstring{\thanks{anastop@upatras.gr}}{}\\
 {\small Laboratory of Universe Sciences, Department of Physics, University of Patras, 26500 Greece} }
\maketitle

\begin{abstract}
We study the time-of-arrival problem for relativistic particles constrained to move on a ring, formulating the problem entirely within Quantum Field Theory (QFT). In contrast to its counterpart for motion in a line, the circle topology implies that particles may encounter the detector multiple times before detection, making a field-theoretic treatment of the measurement interaction essential.
We employ the Quantum Temporal Probabilities (QTP) method to derive a class of Positive-Operator-Valued Measures (POVMs) for time-of-arrival observables directly from QFT. 
We analyze the resulting detection probabilities in both semiclassical and fully quantum regimes, identifying the relevant timescales and their dependence on the field-theoretic parameters. For ensembles of particles, the detection signal is a periodic function, providing a realization of a quantum clock whose operation reflects the local spacetime properties.
We also extend the formalism to rotating rings and show that rotation induces additional noise in detection probabilities, interpretable as a manifestation of the rotational Unruh effect. Finally, we investigate multi-time measurements and demonstrate the emergence of non-classical temporal correlations due to entanglement.
\end{abstract}

\section{Introduction}

The time-of-arrival problem, in its simplest form, may be stated as follows.
A particle is prepared in an initial state $|\psi_0 \rangle$ localized around $x = 0$ and possessing positive mean momentum. If a detector is placed at $x = L$, what is the probability $P(L,t),dt$ that the particle is detected at $x = L$ during the time interval $[t, t+\delta t]$?
The difficulty lies in the fact that this problem admits no unique answer, primarily because a self-adjoint time operator does not exist. Despite its apparent simplicity, there is no consensus on its resolution \cite{ML, ToAbooks}. A variety of approaches have been proposed, relying on markedly different conceptual frameworks.

In this paper, we revisit the time-of-arrival problem in a different setting, namely, that of a particle constrained to move on a ring. This modification substantially alters the character of the problem. In particular, if the detector has a non-zero probability of failing to register the particle upon a single passage, multiple revolutions around the ring may be required before a detection event occurs.

We construct time-of-arrival probabilities for relativistic particles on a ring using the Quantum Temporal Probabilities (QTP) method \cite{QTP1, QTP3, QTP4}. Originally introduced as an algorithm for defining quantum probabilities associated with temporal observables \cite{AnSav06}, the QTP framework has evolved into a general theory of measurements in Quantum Field Theory (QFT)—a subject of increasing foundational and practical significance \cite{QTP1, OkOz, QTP3, FeVe, FJR22, GGM22, PTM, PRA24, QTP4, FeVe2}.

 The central idea underlying QTP is the distinction between the time parameter appearing in the Schrödinger equation and the time variables associated with detection events \cite{Sav99, Sav10}. The latter are treated as macroscopic, quasi-classical variables characterizing the detector. Thus, while the detector is described microscopically by quantum theory, its macroscopic records are expressed in terms of classical spacetime coordinates.

Our motivation for studying the time-of-arrival problem on a ring is threefold.

First, the problem is of intrinsic interest. Many existing approaches to time-of-arrival fail in compact spaces, where particles can encounter the detector multiple times. In such cases, only methods that explicitly incorporate the measurement interaction remain applicable.

Second, the motion of quantum particles on a ring subject to time-of-arrival measurements provides a natural model of a quantum clock. For a large ensemble of identically prepared particles, the detection signal becomes periodic, exhibiting well-defined ''ticks'' that enable precise time tracking. Since our model is formulated within QFT, the clock readings reflect the local structure of spacetime. It is therefore suitable for scenarios in which quantum effects—such as superposition and entanglement—play a significant role \cite{clock, clock2, clock3}.

Third, this set-up offers the simplest framework for investigating the effects of rotation on quantum fields, particularly in relation to their correlation and information-theoretic properties. We find that understanding particle detection on a rotating ring constitutes an important intermediate step toward the study of measurements in rotating frames, with potential applications ranging from laboratory systems \cite{MechanicRotation, MechanicRotation2, experimental observation} to more exotic contexts, such as rotating black holes \cite{Mann, Louko}.

Our main results are as follows:
\begin{enumerate}[i.]
\item We construct explicitly a broad class of Positive-Operator-Valued Measures (POVMs) for the time-of-arrival of relativistic particles on a ring using the QTP method.
Then, we connect these POVMs with their counterparts for particles propagating on the line.

\item We develop an interpretation of this system as a quantum clock and analyze its physical implications.

\item We investigate both the semiclassical and fully quantum regimes of time-of-arrival, identifying the relevant timescales.

\item We extend the construction to rotating rings and show that the rotational Unruh effect \cite{Letaw81, Davies, Bell1, Bell2, Unruh, Levin, Moustos} manifests as an increase in the background noise associated with false detections.

\item We analyze multi-time measurements, highlighting the roles of entanglement and genuinely non-classical features in quantum clock behavior.
\end{enumerate}
 
The structure of the paper is as follows. In Section \ref{section2}, we review prior work, with emphasis on the QTP approach to relativistic measurements and time-of-arrival. In Section \ref{section3}, we construct the POVM for time-of-arrival on a ring and analyze its properties. Section \ref{section4} examines the interpretation of this system as a quantum clock. In Section \ref{section5}, we generalize the analysis to rotating rings, while in Section \ref{section6} we extend the framework to multi-time measurements and entanglement effects. Finally, in Section \ref{section7}, we summarize and discuss our results.

\section{Background}\label{section2}
\subsection{QTP measurement theory}
In this section, we present the main points of the time-of-arrival treatment within the QTP formalism, as presented in Ref. \cite{QTP3}.

QTP emphasizes that every measurement event is localized in spacetime, and that the time of its occurrence is a random variable. This contrasts the von Neumann description of quantum measurements, in which the time of a measurement event is {\em a priori} fixed. The QTP treatment is closer to   experimental practice. Actual particle detectors are fixed in space and the timing of their recordings varies probabilistically. In QTP, the random variables for each detection event include the spacetime point of detection $x$.

Here, we analyse measurements of a scalar field $\hat{\phi}(x)$ on a Hilbert space ${\cal F}$, interacting with an apparatus described by a Hilbert space ${\cal H}$.  
The field and the apparatus interact through a coupling term $\int d^4x \hat{C}(x) \otimes \hat{J}(x) $ (in the interaction picture), where $\hat{C}(x)$ is a local composite operator for the field and $\hat{J}(x)$ is a current operator on the detector.

Assuming an initial state $|\psi\rangle$ for the field and an initial state $|\Omega\rangle$ for the detector,
 the unnormalised probability density $P(x)$ for a detector record at a spacetime point $x$ is,
\bey
P(x) = C \int d^4 y    R(y) G(x - \frac{1}{2}y, x +\frac{1}{2}y), \label{prob1aa}
\eey
to leading order in the interaction. In Eq. (\ref{prob1aa}), $C$ is a normalization constant
\bey
G(x, x')  = \langle \psi|\hat{C}(x) \hat{C}(x')|\psi\rangle
\eey
is a two-point correlation function for the field, and
\bey
R(x) := \langle \Omega| \hat{J}(x_0)  e^{-i\hat{P}\cdot x}\hat{J}(x_0)|\Omega\rangle \label{detkern}
\eey
is the {\em detection kernel}. In Eq. (\ref{detkern})  $x_0$ is a reference point in the detector's worldline,  and $\hat{P}_{\mu}$ is the detector's energy momentum operator.

Eq. (\ref{prob1aa}) can be derived simply by mimicking spacetime sampling with a switching function in the Hamiltonian, or in a conceptually more satisfying way, by analysing spacetime sampling on the detector using a decoherent histories analysis of measurement records. The two analyses yield the same results to leading order in the field-apparatus coupling, but they differ in higher orders.

  The expressions for the detector kernel simplify under the  assumptions that the detector (i)   follows an inertial trajectory in Minkowski spacetime, and (ii) is initially prepared on a state $|\Omega\rangle$ that is   approximately translation invariant.
 Then, the detector kernel takes the form $R(x) = \langle w|e^{-i \hat{P} \cdot (x - x_0)}|w\rangle$.

Its Fourier transform $\tilde{R}(\xi) = \int d^4x e^{-i \xi\cdot x} R(x)$  
is given by
\bey
\tilde{R}(\xi) = (2\pi)^4 e^{i \xi\cdot x_0} \langle w|\hat{E}_{\xi}|w\rangle \geq 0, \label{tildr}
\eey
where $\hat{E}_{\xi} = \delta^4(\hat{p} - \xi)$ is the projector onto the subspace with four-momentum $\xi^{\mu}$.
The momentum four vector associated with the detector is timelike and the associated energy $p^0$ is positive. This means that $\tilde{R}(p) = 0$ for spacelike $p$, or if $p^0 < 0$.  

\subsection{Time-of-arrival measurements in the real line}

We specialize to   time-of-arrival measurements for a   free scalar field  of mass $\mu$,
\bey
\hat{\phi}(x) = \int \frac{d^3k}{(2\pi)^{3/2} \sqrt{2\omega_{\bf k}}} [\hat{a}_{\bf k} e^{-ik\cdot x} + \hat{a}^{\dagger}_{\bf k} e^{ik\cdot x}],
\eey
expressed in terms of annihilation operators $\hat{a}_{\bf k}$ and creation operators $\hat{a}^{\dagger}_{\bf k}$; we  have set  $k = (\omega_{\bf k}, {\bf k})$ with $\omega_{\bf k} = \sqrt{{\bf k}^2 + \mu^2}$.

  We assume that the source is localized around ${\bf x} = 0$, and that there is a single detector at ${\bf x}$; both the source and the detector  are motionless.  

If the size of the detector is much smaller than the source-detector distance, then particles effectively propagate in one dimension, along the line from the source to the detector. Hence, we will express each spacetime coordinate as a pair $(t, x)$, where $t$ is the time in the rest-frame where the detector is static and $x$ is the distance from the source. It is convenient to take the positions of the detection events as fixed and treat only detection times as random variables.

Hence, the probability densities for detection at $x > 0$  becomes
\bey
P(t, x) &=&  C \int \frac{dk}{2\pi \sqrt{2\omega_k}} \frac{dk'}{\sqrt{2\omega_k'}}  \rho_1(k, k') \tilde{R}\left(\frac{1}{2}(k+k'), \frac{1}{2}(\omega_k+\omega_{k'}) \right)\nonumber \\
&\times& e^{i(k-k')x - i (\omega_k - \omega_{k'})t}, \label{p1t1} 
\eey
where $\rho_1(k, k') = \langle \psi|\hat{a}^{\dagger}_{k'} \hat{a}_k|\psi\rangle$ is the one-particle reduced density matrix for the field state $|\psi\rangle$.
Note that here the quantities $k$ and $k'$ represent spatial momenta, and not four-momenta.

 The detection probability for negative momentum particles is negligible for sufficiently large source-detector distances. It is therefore convenient to assume that the density matrices have support only on positive momenta.  In this case there is a simple normalization of those probabilities when taking time to range in the full real axis,
 \bey
 \int_{-\infty}^{\infty} dt P(t,x) = C \int \frac{dk}{2k} \tilde{R}(k, \omega_k) \rho_1(k,k).
 \eey
 The quantity $\alpha(k):= \tilde{R}(k, \omega_k)/(2k)$ is the {\em absorption coefficient} of the detector, which gives the fraction of incoming particles of momentum $k$ that are absorbed.  
 We can choose $C$  so that  $\int_{-\infty}^{\infty} dt P(t, x)  = 1$. 
 We also redefine density matrices, post-selected along detection,
 \bey
 \tilde{\rho}_1(k,k') = C_1 \rho_1(k,k') \sqrt{\alpha(k)\alpha(k')} \label{ro1redef}
 \eey
 Then,
\bey
P(t, x)  = \int \frac{dk dk'}{2\pi} \tilde{\rho}_1(k,k') \sqrt{v_k v_{k'}}  L(k,k')  e^{i(k-k')x - i (\omega_k - \omega_{k'})t}, \label{p1tt}
\eey
 where $v_k = k / \omega_k$ is the relativistic velocity,
  and
 \bey
L(k, k') = \frac{\tilde{R}\left(\frac{1}{2}(k+k'), \frac{1}{2}(\omega_k+\omega_{k'}) \right)}{\sqrt{\tilde{R}(k, \omega_k) \tilde{R}(k', \omega_{k'})}}
 \eey
 are the matrix elements of the {\em localization operator} $\hat{L}$. This operator is defined on the Hilbert space ${\cal H}_1$ of a single particle, and it determines the position spread of the detection record in the apparatus \cite{QTP3}. By construction, $L(k, k) = 1$. Positivity of probabilities implies that $L(k, k') \leq 1$.
 
 It can be shown that maximum localization is achieved for  $L(k, k') = 1$, in which case $\hat{L} = \delta (\hat{x}_{NW})$, where  $\hat{x}_{NW}$ is the Newton-Wigner position operator \cite{NWi}. 
 Maximum localization corresponds to 
  an exponential form for the detection kernel ${\tilde R}(k, \omega)$,
\bey
{\tilde R}(k, \omega) = \left\{ \begin{array}{cc} A \exp \left(-\gamma_1 |k| - \gamma_0 \omega   \right), & |k|\leq \omega,
\\
0& |k| > \omega, \; \mbox{or} \; \omega <0, \end{array}\right.
\eey
for some constants $\gamma_1, \gamma_0 \geq 0$, and $A > 0$.

We can also write $P(t, x)  =  Tr[\hat{\tilde{\rho}}_1 \hat{\Pi}_{t, x} ]$
where the POVM $\hat{\Pi}_{t, x} $ is defined by
\bey
\hat{\Pi}_{t, x}  = \sqrt{\hat{v}} \hat{U}^{\dagger}(x, t) \hat{L}\hat{U}(x, t)  \sqrt{\hat{v}}. \label{pxtt}
\eey
For maximum localization, $\langle x'| \hat{\Pi}_{t, x}  |x''\rangle = {\cal A}^{(0)*}_t(x'-x) {\cal A}^{(0)}_t(x''-x)$, where
\bey
 {\cal A}_t^{(0)}(x) = \int dk \sqrt{v_k}e^{ikx - i \omega_kt}. \label{ampllk}
\eey
This  POVM    coincides with the Leon-Kijowski POVM for the time of arrival of relativistic particles \cite{Leon, Kij}.

\section{Time-of-arrival probabilities in a ring}\label{section3}

In this section, we will derive the Time-of-Arrival Probabilities for particles moving in a ring. The derivation follows the QTP prescription, as described in Section \ref{section2}. The key difference lies in the fact that a ring is a compact manifold, so that the particle will have multiple opportunities to be detected if it does not succeed in the first one. 

\subsection{Scalar field theory on a (1+1) dimensional cylinder}

To describe time-of-arrival measurements in a ring, we employ the 
  quantization of scalar fields in the Minkowski spacetime using cylindrical coordinates, which has been studied extensively \cite{Letaw, Ambrus}. 
  
  Since time-of-arrival measurements essentially restrict the system to one spatial dimension, the analysis is equivalent to the analysis of a scalar field in a cylindrical two-dimensional spacetime 
  $\mathbb{R}  \cross \mathbb{S}^1$. 
  The Klein-Gordon equation for a real scalar field $\phi$ is then given by:

\begin{equation}
    \left[-\partial^2_t+\frac{1}{r^2}\partial^2_\varphi-\mu^2\right]\hat{\phi}(\varphi,t)=0,
\end{equation}
where $\mu$ is the mass of the field and $\varphi$ the angular variable. The radius $r$ of the cylindrical coordinates is a free parameter. The mode expansion of the field is:
\begin{equation}
    \hat{\phi}(\varphi,t)=\sum_m\frac{1}{\sqrt{8\pi^2r\omega_m}}\left(\hat{a}_{m}e^{-i\omega_m t+im\varphi}+\hat{a}^\dagger_{m}e^{i\omega_m t-im\varphi}\right), \label{minkphi}
\end{equation}
where $\hat{a}_m$ and $\hat{a}_m^{\dagger}$ are annihilation and creation operators respectively. We also wrote, 
  $\omega_m^2=\mu^2+\frac{m^2}{r^2}$. 

The single-particle Hilbert space is $L^2(S^1, r  d\varphi)$. Physical self-adjoint operators include the angular momentum operator $\hat{\ell} = - i \frac{\partial}{\partial \varphi}$ and the trigonometric operators $\hat{c}$ and $\hat{s}$, defined by $\hat{c} \psi(\varphi) = \cos\varphi \psi(\varphi) $ and $\hat{s} \psi(\varphi) = \sin\varphi \psi(\varphi) $. 
The operators $\hat{\ell}$, $\hat{c}$, and $\hat{s}$ define a unitary representation of the Euclidean group $E_2$.

The standard basis on $L^2(S^1, r  d\varphi)$ is defined by eigenvectors $|m\rangle$  of the angular momentum operator, $\hat{\ell}\ket{m} = m \ket{m}$.
The Hamiltonian operator is $\hat{h} = \sqrt{\mu^2 + r^{-2} \hat{\ell}^2}$.

\subsection{Time-of-arrival probability density}

The general probability density for Time-of-Arrival measurements is given by \cite{QTP3}:
\begin{equation}\label{eq3}
     P(t,\varphi)= C \int ds\int dy R(s,y)G(t-s/2,\varphi-y/2;t+s/2,\varphi+y/2),
\end{equation}
where $R(t,\varphi)$ is the detector kernel, encoding the physical properties of the measurement apparatus and $G(t_1,\varphi_1;t_2,\varphi_2)$ is the Wightman correlation function for the field. In our case, equation \eqref{eq3} takes the form:
\begin{equation}\label{eq4}
   \begin{split}
        & P(t,\varphi)=  P_0 +\\
        &C \sum_{m,m'}\frac{1}{4\pi r\sqrt{\omega_m\omega_{m'}}}\rho(m,m')\tilde{R}\left(\frac{\omega_m+\omega_{m'}}{2},\frac{m+m'}{2}\right)e^{i(m-m')\varphi-i(\omega_m-\omega_{m'})t},
   \end{split}
\end{equation}
where $\tilde{R}(\omega_m,m)$ is the Fourier transform of $R(t,\varphi)$ and $\rho(m,m')=\bra{\psi}\hat{a}^\dagger_{m'}\hat{a}_{m}\ket{\psi}$ is the single-particle reduced density matrix. The first term in the right-hand side:

\begin{equation}
     P_0 = C \sum_m\tilde{R}(\omega_m ,m)\frac{1}{4\pi r\omega_m},
\end{equation}
is state-independent and corresponds to the vacuum noise. Therefore, it can be omitted from \eqref{eq4}.

Note that the derivation of Eq. (\ref{eq4}) assumes that the initial state is localized close to the source. This means that the state receives no contribution from the zero modes: $\rho(0, 0) = 0$.

\subsection{Normalization and post-selection of detection events}\label{sec1.3}

It is convenient to select a normalization condition, in order to fix the multiplicative constant $C$. We saw that for a particle on a line, the condition $\int_{-\infty}^{\infty} dt P(t, x) = 1$ for initial states with positive momentum is convenient, and it leads to a natural definition of the absorption coefficient and the localization operator.

Unfortunately, this condition cannot be applied to a particle moving in a ring, because the integral $\int_{-\infty}^{\infty} dt P(t, \varphi)$ diverges for discrete momentum. 
A physical normalization condition involves integration in a finite time interval $[0, T]$, leading to a $T$ dependence of the normalization constant. However, it is more convenient to regularize the probability density $P(t)$ in order to define a finite integral. The normalization, although not physical, allows us to study the structure of the time-of-arrival probability on the circle. For an actual experiment of duration $T$, we can always normalise to unity by numerically intergrating $\int_0^T dt P(t,\varphi)$.
\color{black}Since $P(t)$ is of the form $\sum_{m,m'}G(m,m') e^{-i(\omega_m - \omega_{m'})t}$ for some function $G(m, m')$, we define the regularized version as
\bey
P_{\gamma}(t) = \int dy \int dy' \sum_{m,m'} f_{\gamma}(y - m) f_{\gamma}(y'-m') 
G(y,y')e^{-i(\omega_y - \omega_{y'})t},
\eey
where $f_{\gamma}$ is an approximate delta function, i.e., a family of functions labeled by $\gamma$, such that $\lim_{\gamma\rightarrow 0}f_{\gamma}(y) = \delta (y)$. For example, we can take a Gaussian $f_{\gamma}(y) = (\pi \gamma^2)^{-1/2} \exp\left(-y^2/\gamma^2\right)$. Then, the total probability becomes
\bey
P_{\gamma} =  \int_{-\infty}^{\infty} dt P_{\gamma}(t)= 2 \pi \int dy   \sum_{m,m'} f_{\gamma}(y - m) f_{\gamma}(y-m')  F(y, y')  \frac{r^2 \omega_y}{y}.
\eey
At the limit $\gamma \rightarrow 0$, $f_{\gamma}(y - m) f_{\gamma}(y-m') \simeq \delta(y - m) \delta_{mm'}/ B(\gamma)$, where $B(\gamma)$   depends on the family of functions $f_{\gamma}$. (For the Gaussians $f_{\gamma}$, $B(\gamma) = \sqrt{\pi}\gamma  $.) Then, we obtain
\bey
P_{\gamma} = \frac{2\pi r^2}{B(\gamma)} \sum_m F(m, m) \frac{\omega_m}{m} = \frac{C }{B(\gamma)} \sum_m \rho(m,m)\frac{r}{2m }\tilde{R}(\omega_m,m). \label{pgamma}
\eey
Define $a(m)= \frac{C r}{B(\gamma)} \frac{\tilde{R}(\omega_m,m)}{2m }$, an analogue of the absorption coefficient of the detector. This quantity   determines  the fraction of  particles with angular momentum $m$ that are detected.

We define a normalized conditional probability distribution $P_c(t,\varphi)=P(t,\varphi)/P_{\gamma}$, 

\begin{equation}\label{eq7}
     P_c(t,\varphi)= B(\gamma)\sum_{m,m'}\frac{\rho_{ps}(m,m')}{2\pi r} \bra{m}\hat{L}\ket{m'}\sqrt{v_m v_{m'}}e^{i(m-m')\varphi-i(\omega_m-\omega_{m'})t},
\end{equation}
 where $v_m =   m/(\omega_mr)$ is the {\em linear velocity} of the particle,
 and 
 \bey
 \rho_{ps}(m,m')=\frac{\rho(m,m')\sqrt{a(m)a(m')}}{P_{\gamma}} \label{reedef}
 \eey
 is the  density matrix post-selected for the different detection rates at different $m$.
 
  We have defined the localization operator   
  \bey
  L(m,m') = \bra{m}\hat{L}\ket{m'} = \frac{\tilde{R}\left(\frac{\omega_m+\omega_{m'}}{2},\frac{m+m'}{2}\right)}{\sqrt{\tilde{R}\left(\omega,m\right)\tilde{R}\left(\omega',m'\right)}}. \label{locop}
  \eey
  The localization operator is positive by construction, has positive matrix elements in the momentum basis and satisfies $L(m, m) = 1$ for all $m$.

  Eq. (\ref{eq7}) can be written as
  \bey
 P_c(t,\varphi)= \frac{B(\gamma)}{2\pi r} Tr \left(  e^{-i\hat{Ht} +i \hat{\ell}\varphi} \sqrt{\hat{v}}\hat{\rho}_{ps}\sqrt{\hat{v}} e^{i\hat{Ht} -i \hat{\ell}\varphi}  \hat{L}\right),
 \label{pcdd}
  \eey
which takes the form $\frac{B(\gamma)}{2\pi r} Tr[\hat{\rho}_{ps}\hat{\Pi}_{t, \varphi}]$, expressed in terms of positive operators
\bey
\hat{\Pi}_{t, \varphi} =   \sqrt{|\hat{v}|}e^{i\hat{Ht} -i \hat{\ell}\varphi}\hat{L} e^{-i\hat{Ht} +i \hat{\ell}\varphi} \sqrt{|\hat{v}|}. \label{ptphi}
\eey

\subsection{Localization operator}
We will analyze the structure of the time-of-arrival probabilities, using the Wigner-Weyl representation for a particle in a ring \cite{Wigner}. For each operator $\hat{A}$ on the single-particle Hilbert space $L^2(S^1, r d\varphi)$ we define a function $\tilde{A}(\theta, p)$ on the classical phase space $\Gamma =   \{ (\theta,p)\in \mathbb{R}^1 \cross \mathbb{S}^1 \}$,

\begin{equation}
    \tilde{A}(\theta,p)=\frac{1}{2\pi}\sum_{m,m'}\langle m|\hat{A}|m'\rangle e^{i(m'-m)\theta}\sinc{\pi(p-(m+m')/2)}.
\end{equation}
The sinc function interpolates between the discrete angular momentum $m$ and the continuous variable $p$. 

Since   sinc is a positive definite function, the Wigner-Weyl transform $\tilde{L}(\theta,p)$ of the localization operator $\tilde{L}(\theta, p) \geq 0$. This follows from the Bochner–Herglotz theorem and the fact that $L(m, m') \geq 0$. We also find that $\int d\theta \tilde{L}(\theta,p)  = 1$, which implies that 
 $\tilde{L}(\theta,p)$ is a probability distribution with respect to $\theta$, parameterized by $p$. It defines the irreducible spread of the measurement record on the ring $S^1$ for a given momentum $p$.

Since $\hat{L}$ is a positive operator, the Cauchy-Schwarz inequality applies: $L(m,m') \leq \sqrt{L(m,m) L(m',m')} = 1$. The Wigner-Weyl transform of the localization operator with $L(m, m') = 1$  is
\bey
\tilde{L}(\theta, p) = \delta(\theta),
\eey
hence, this operator corresponds to maximum localization \cite{QTP3}.

\subsection{The POVM of maximum localization}\label{sec.3.5}
For maximum localization, the positive operators $\tilde{\Pi}_{t, \varphi}$ factorize, 
\begin{equation}\label{Piop}
    \bra{m}\hat{\Pi}_{t,\varphi}\ket{m'}=\sqrt{v_mv_{m'}}e^{i(m-m')\varphi-i(\omega_m-\omega_{m'})t},
\end{equation}
or, in position space $\bra{\theta}\hat{\Pi}_{t,\varphi}\ket{\theta'} = {\cal A}_t(\varphi - \theta  ) {\cal A}_t(\varphi - \theta') $,
in terms of the amplitude
\bey
{\cal A}_t(\varphi) = \sum_{m = 0}^{\infty} \sqrt{|v_m|}e^{im\varphi - i \omega_m t} = \sum_{m = -\infty}^{\infty} \Theta(m) \sqrt{|v_m|}e^{im\varphi - i \omega_m t}, \label{atphi}
\eey
where $\Theta$ stands for the step-function.

Using the Poisson expansion $\sum_{m=-\infty}^{\infty} f(m) = \sum_{n=-\infty}^{\infty} \int_{-\infty}^{\infty} dx f(x)e^{i2\pi nx}$, we can write the amplitude (\ref{atphi}) 
as a series 
\bey
{\cal A}_t(\varphi) =  \sum_{n = -\infty}^{\infty} r{\cal A}^{(0)}_t[(\varphi + 2\pi n)r] \label{atat0}
\eey
where ${\cal A}^{(0)}_t(x)$ is the amplitude (\ref{ampllk})
 corresponding to the Leon-Kijowski POVM in the real line.

The amplitude ${\cal A}^{(0)}_t(x)$ is straightforwardly evaluated in the saddle-point approximation, and we find that 
\bey
{\cal A}_t(\varphi) = \sum_{n >0} \theta [t - (\varphi +2\pi n)r]\frac{\sqrt{2\pi i \mu r^3 t(\varphi +2\pi n)}}{[t^2 - (\varphi +2\pi n)^2 r^2 ]^{3/4}}
e^{-i \mu \sqrt{t^2-(\varphi +2\pi n)^2 r^2  }}.
\eey
The amplitude exhibits an obvious periodicity, and it is causal, as it vanishes outside the light-cone. 
For $\mu = 0$, the integral is evaluated exactly, and ${\cal A}_t(\varphi) =  \sum_{n > 0} \delta(\varphi + 2 \pi n - t)$.

The simple relation (\ref{atat0}) does not translate into the Wigner-Weyl transform of 
   $\hat{\Pi}_{t,\varphi}$. To understand its phase space structure, it is more convenient to use the Q-symbol. The Q-symbol of any operator $\hat{A}$ is a function $Q_A(\theta, p)$ on the state space $\R\times \mathbb{S}^1$, defined as
   \bey
Q_A(\theta, \xi) = \langle \theta, \xi|\hat{A}|\theta, \xi\rangle
   \eey
in terms of the generalized coherent states of the group $E_2$ that describe a particle in a ring,
\bey
|\theta, \xi\rangle = C_{\xi} \sum_{m=-\infty}^{\infty} e^{-\frac{1}{2\alpha^2}(m - \xi)^2 - i m \theta}|m\rangle, \label{cohs}
\eey
where the normalization constant
\bey
C_{\xi} = \frac{1}{(\pi\alpha^2)^{1/4}}   [\theta_3(-\pi \xi,e^{-\pi^2 \alpha^2})]^{-1/2}
\eey
is expressed in terms of the Jacobi theta-function
\cite{Abramovich}
\bey
\theta_3(x, y) = 1 + 2 \sum_{n=1}^{\infty}y^{n^2} \cos(2nx).
\eey
The coherent states (\ref{cohs}) generalize those of Ref. \cite{KRP}---the latter are obtained from Eq. (\ref{cohs}) for $\alpha = 1$. They are peaked around $m = \xi $ with a spread of order $\alpha$. For $\xi > 0$, the contribution of negative momenta to the coherent states is suppressed by terms of order $e^{-\xi^2 \alpha^2}$, so they are well compatible with the restriction to positive values of $m$.

   We compute the Q-symbol of $\hat{\Pi}_{t,\varphi}$
\bey
Q_{\Pi_{t, \varphi}}(\theta, \xi) = |{\cal B}_t(\varphi- \theta, \xi)|^2, \label{qsymbol}
\eey
where
\bey
{\cal B}_t(\varphi, \xi) =   C_{\xi}\sum_m \sqrt{v_m} e^{im\varphi - i \omega_m t  -\frac{1}{2\alpha^2}(m - \xi)^2}.  \label{btpx}
\eey

Again, using the Poisson expansion method, we write
\bey
{\cal B}_t(\varphi) = c_{\xi} \sum_{n=-\infty}^{\infty}  {\cal B}_t^{(0)}(r(\varphi + 2 \pi n), \xi r^{-1}),\label{cxi}
\eey
where the amplitude for time-of-arrival in the real line
\bey
{\cal B}_t^{(0)}(x, p) = \int_0^{\infty} dk \sqrt{v_k} e^{ikx -i \omega_kt} \tilde{\psi}_{0,p}(k) =\sqrt{2\pi} \int dx' {\cal A}_t^{(0)}(x-x')\psi_{0,p}(x')
  \eey 
is expressed in terms of an initial Gaussian coherent state 
\bey
\psi_{0,p}(x) = (2\pi \sigma^2)^{-1/4} e^{-\frac{x^2}{4\sigma^2} +ipx}, \label{gacoh}
\eey
localized around $x = 0$, with mean momentum $p$, and position spread  $\sigma = \frac{1}{\sqrt{2}} r/\alpha$. In Eq. (\ref{cxi}), $c_{\xi} = [\theta_3(-\pi \xi,e^{-\pi^2 \alpha^2})]^{-1/2}$.

For a well-defined time-of-arrival measurement there must be negligible overlap between the initial state and the detector, so we work in the regime $\alpha >> 1$. In this regime, $c_\xi $ is well approximated by unity, the leading order correction being of   order   $e^{-\pi^2 \alpha^2}$ \cite{Abramovich}.

For $p \sigma >> 1$, $ {\cal B}_t^{(0)}(x, p) \simeq \sqrt v_p \psi_{0,p}(x, t)$ where $\psi_{0,p}(x, t)$ is the time evolution of the Gaussian coherent state (\ref{gacoh}). This is a wave-packet peaked around $x = v_p t$, with spread $\sigma^2(t) = \sigma^2 +(\Delta v)^2 t^2$, where $\Delta v$ is the mean deviation of the velocity operator $\hat{v} = \hat{p}\hat{H}^{-1}$ on the state (\ref{gacoh}).

We find that to leading order in $(p\sigma)^{-1}$, $(\Delta v)^2 = \frac{\mu^4}{2\epsilon_p^6\sigma^2}$, so that 
\bey
\sigma^2(t) = \sigma^2 + \frac{\mu^4}{2\epsilon_p^6\sigma^2} t^2.
\eey
As long as $\sigma(t) << r$, the partial amplitudes contributing to ${\cal B}_t(\varphi, \xi)$ by Eq. (\ref{cxi}) have no overlap, hence, 
\bey
Q_{\Pi_{t, \varphi}}(\theta, \xi) = \sum_{n=-\infty}^{\infty} |{\cal B}_t^{(0)}(r(\varphi - \theta + 2 \pi n),  \xi r^{-1})|^2. \label{qpqp}
\eey
The Q-symbol has peaks at $t_n =  r(\varphi - \theta + 2 \pi n)/v_p$, for $p = \xi/r$,   with width $\sigma(t_n)$, for positive $n$, as long as  $\sigma(t_n) << r$. In this regime, $n$ can be interpreted as the winding number of a classical path in the circle. 

For $\mu = 0$, Eq. (\ref{qpqp}) is accurate at all times, because the spread $\sigma(t)$ remains constant. 
However, for $\mu \neq 0$, when the winding number becomes of the order of 
  $\omega_{\xi}^2 p \sigma/\mu^2 $, 
the approximation in Eq. 
(\ref{qpqp}) fails, as the superposition of partial amplitudes at different $n$ becomes significant. Hence, there is a ``quantum" time-scale \cite{Nauenberg}
\bey
T_q = \frac{\omega_{\xi}^3 r^2}{\mu^2 \alpha},
\eey
 at which the semiclassical approximation breaks down.

\begin{figure}
\centering

  \begin{subfigure}{.4 \textwidth}
  \caption{Semiclassical regime, $t \simeq 0.07 \, T_q$.}
    \includegraphics[width=\linewidth]{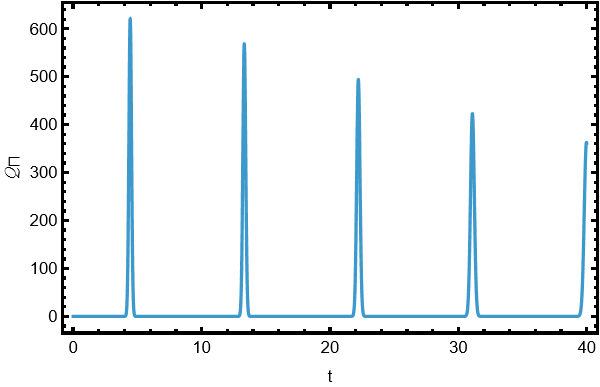}
  \end{subfigure}%
  \begin{subfigure}{.4 \textwidth}
   \caption{Semiclassical regime, $t \simeq 0.6 \, T_q$.}
    \includegraphics[width=\linewidth]{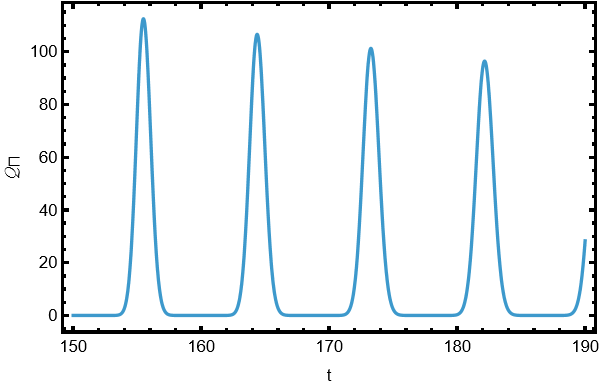}
  \end{subfigure} 
  \begin{subfigure}{.4\textwidth}
  \caption{End of semiclassical regime, $t \simeq  1.5 T_q$}
    \includegraphics[width=1\linewidth]{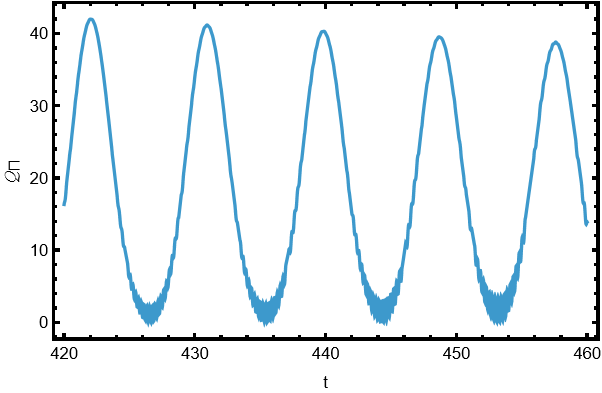}
  \end{subfigure}%
  \begin{subfigure}{.4\textwidth}
  \caption{Onset of quantum regime, $t \simeq 1.8 \, T_q$.}
    \includegraphics[width=1\linewidth]{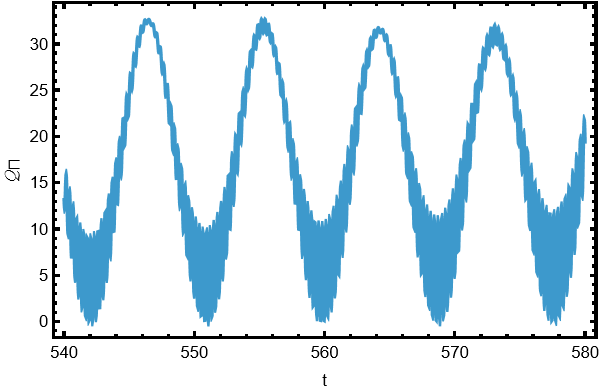}
  \end{subfigure}
   \begin{subfigure}{.4\textwidth}
     \caption{Deep quantum regime, $t \simeq 4.2\, T_q \sim 0.03 T_{rec}$.}
    \includegraphics[width=1\linewidth]{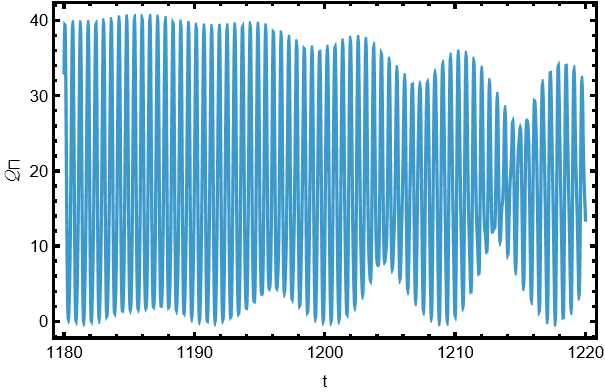}
  \end{subfigure}%
  \begin{subfigure}{.4\textwidth}
   \caption{Deep quantum regime, $t \simeq 10 \, T_q \sim 0.08 T_{rec}$.}
    \includegraphics[width=1\linewidth]{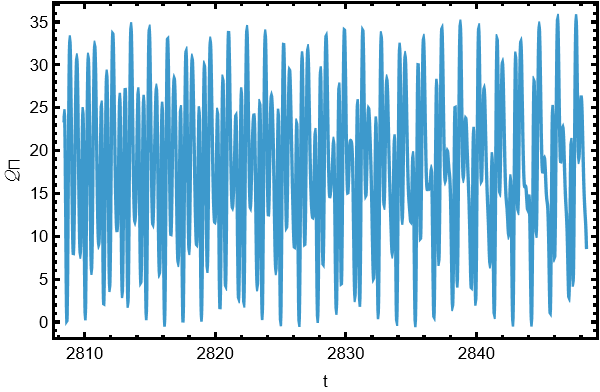}
  \end{subfigure}
   \begin{subfigure}{.4\textwidth}
      \caption{Partial revival, $t \sim 0.98 T_{rec}$. }
    \includegraphics[width=1\linewidth]{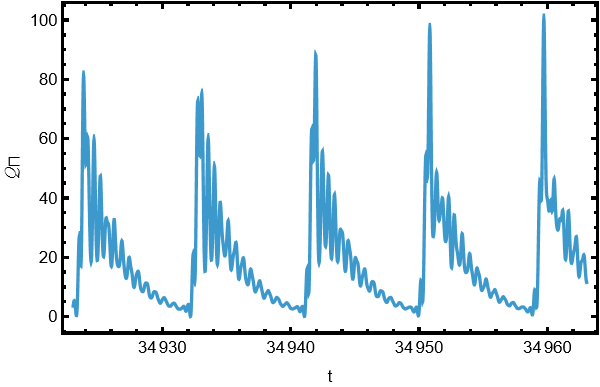}
  \end{subfigure}%
  \begin{subfigure}{.4\textwidth}
   \caption{Partial revival, $t \sim   T_{rec}$. }
    \includegraphics[width=1\linewidth]{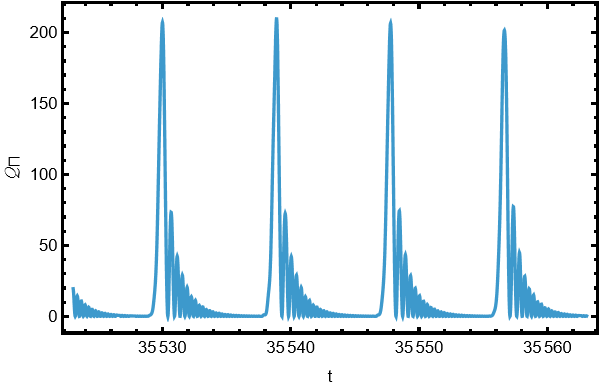}
  \end{subfigure}
  
\caption{The $Q$-symbol (\ref{qsymbol}) for $\varphi = \pi$, $\xi=1000, \mu=1000$, $r = 1$, $\alpha = 10$ at different time-scales where $T_q\approx282$ and $T_{rev}\approx35500$.}
\label{probcoh}
\end{figure}

This breakdown does not last indefinitely. The spectrum of the Hamiltonian is discrete, so the probability amplitude is subject to the {\em quantum recurrence theorem}
\cite{Quantum Recurrence Theorem}. There is a recurrence time-scale $T_{rec}$ that characterizes wave-packet revivals, estimated as   $T_{rec} \sim 2\pi |\omega_{\xi}''|^{-1}$, where the prime denotes derivative with respect to $m$ \cite{wavepacket revivals}. We find that 
\bey
T_{rec} = 4\pi \frac{\omega_{\xi}^3 r^2}{\mu^2} = 4 \pi\alpha  T_q.
\eey
We can see both the breakdown of the semi-classical approximation and the partial recurrence in Fig. \ref{probcoh}.

Since the coherent states (\ref{cohs}) form an overcomplete family of states, we can express most states $\hat{\rho}$ in terms of their 
$P$-symbol $P_{\rho}(\theta, \xi)$ defined by 
\bey
\hat{\rho} = \int d \theta d \xi P_{\rho}(\theta, \xi) |\theta, \xi\rangle \langle \theta, \xi|.
\eey
Then, the time-of-arrival probabilities are   expressed as
\bey
P_c(t, L) = \frac{B(\gamma)}{2\pi r} \int d \theta d \xi P_{\rho}(\theta, \xi) Q_{\Pi_{t, \varphi}}(\theta, \xi),
\eey
and they are in principle calculable if the Q-symbol $Q_{\Pi_{t, \varphi}}(\theta, \xi)$ has been determined.

\subsection{Generalizations}
Using coherent states, it is straightforward to express the time-of-arrival probabilities for a general detector. Denoting the P-symbol of the localization operator by $\hat{P}_S$, we find that
\bey
P_c(t, \varphi) = \frac{B(\gamma)}{2\pi r}  \int d\theta d \xi d\theta' d \xi' P_{\rho}(\theta, \xi)P_{S}(\theta', \xi')|\langle \theta', \xi'|e^{-i\hat{Ht} +i \hat{\ell}\varphi} \sqrt{\hat{v}}|\theta, \xi\rangle|^2.
\eey
It is straightforward to show that, modulo terms of order $e^{-\alpha^2 \pi^2}$,
\bey
\langle \theta', \xi'|e^{-i\hat{Ht} +i \hat{\ell}\varphi} \sqrt{\hat{v}}|\theta, \xi\rangle = \frac{1}{(2\pi \alpha^2)^{1/4}} e^{- \frac{(\xi - \xi')^2}{4\alpha^2}} {\cal B}_t(\varphi - \theta + \theta', \xi+\xi'),
\eey
where the amplitude ${\cal B}_t(\varphi, \xi)$ is given by Eq. (\ref{btpx})---but with spread parameter $\sqrt{2} \alpha$ rather than $\alpha$. The calculations can then proceed in a way similar to Sec. \ref{sec.3.5}.

Our analysis so far has been restricted to initial states with non-negative momenta---modulo corrections that are exponentially suppressed in the coherent states (\ref{cohs}). This leads to a convenient normalization condition with time in the full real axis. There is then a natural definition of the absorption coefficient $\alpha(p)$, of the conditional initial state (\ref{reedef}) and of the localization operator (\ref{locop})---in accordance with the POVMs for the time of arrival in the real line. 

The key point is that the same definitions work also for states with support in negative angular momenta $m$. It is therefore possible to extend those definitions to negative angular momenta $m$, leading to the positive operator (\ref{ptphi}). The full analysis of Secs. 3.4 and 3.5 then passes through with this generalization. The resulting POVM, however, does not satisfy a simple normalization condition:  Eq. (\ref{pgamma}) is spoiled by the presence of interference terms between positive and negative momenta.

\section{Particles in a ring as quantum clocks}\label{section4}

A clock is a physical system that  
possesses an observable quantity 
$q(t)$, whose evolution is approximately periodic 
$q(t+T)=q(t)$ and the period is reproducible under specified initial conditions. Time intervals are then defined 
   by counting the number of "ticks", that is, of completed cycles.

The motion of quantum particles in a ring subject to time-of-arrival measurements does qualify as a clock. Consider $N$ particles ($N>> 1$) prepared in a state that is sharply localized in position and momentum. Every time they pass through the detector, a fraction of those particles is detected, corresponding to the peaks in the probability density $P(t)$ of the time of arrival. Each peak defines a tick of the clock.

\begin{figure}[H]
  \centering
   \includegraphics[scale=0.8]{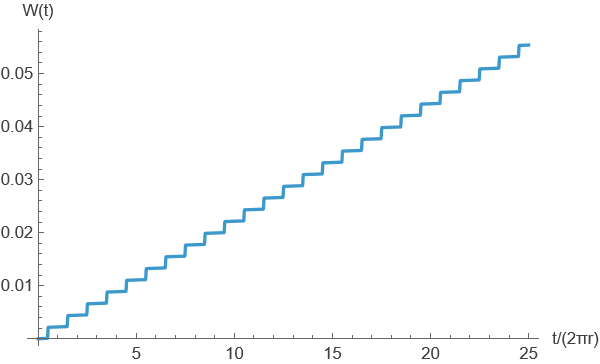}
   \caption{The cumulative probability density $W(t)$ as a function of the dimensionless time $t/(2\pi r)$.}
  \label{steps}
\end{figure}

This is most easily seen by considering the detection probability for massless particles. For an initial coherent state $|0, \xi\rangle$ at the limit $\xi \alpha >> 1$, and for a detector located at coordinate $\varphi$ we find that
\bey
P_c(t, \varphi) = \frac{   B(\gamma)}{2 \pi r} \sum_{n=-\infty}^{\infty} |\psi_{0, \xi r^{-1}}[r(\varphi + 2 \pi n) - t ]|^2,
\eey
in terms of the Gaussian coherent states $\psi_{x,p}$ with width $\sigma = \frac{1}{\sqrt{2}} r/\alpha$. There are sharp peaks at $t = r(\varphi + 2 \pi n) $ with a constant width. The detection signal behaves as a clock, as can be seen in the behavior of the cumulative detection probability at time $t$, $W(t) = \int_0^t ds P_c(s, \varphi) $ in Figure \ref{steps}. A clock tick is a jump in the total number of detected particles and corresponds to a time step of $\tau = 2 \pi r$.


There are several benefits to this clock model. First, it is defined in terms of signals from a fundamentally quantum system, so that it can be used to test the behavior of quantum clocks in presence of quantum phenomena, such as superposition or entanglement. 

Of equal importance is the fact that the associated probabilities are constructed in terms of quantum field interactions, depending essentially on the field two-point function---see Eq. (\ref{prob1aa}). This means that the clock rate captures local properties of spacetime, hence, it can act as a quantum probe of the local geometry.

For example,   a Newtonian gravitational field with potential $\Phi(x)$ acts on the   electromagnetic field  as a dielectric with index of refraction $n(x) = 1 + 2 |\Phi(x)|$ \cite{ir1, ir2, ir3, ir4}. If the dimensions of the ring are much smaller than the typical rate of change of $\Phi$, then two identical rings at locations $x_1$ and $x_2$ would have ratio of periods 
\bey
T_2/T_1 = \frac{1+2 |\Phi(x_2)|}{1 +2 |\Phi(x_1)| },
\eey
in accordance with the gravitational time dilation.

In contrast, if the potential changes significantly along the ring, the behavior of the clock may be strongly modified, possible leading to a higher rate of errors.  

More generally, we may consider the embedding of a ring in a spacelike three-surface of a general spacetime. A sufficiently small ring will capture the behavior of the lapse function of the metric, and, as such, it could be used to investigate horizons or gravitational singularities.  Furthermore, by considering vector fields rather than scalars, we might be able to connect the clock behavior to curvature effects, for example the holonomy of the particle’s polarization around the ring can provide a measure of the Riemann tensor. These possibilities will be explored in a later publication.

We saw that for massless particles and for a maximum-localization apparatus, the clock is ideal, in the sense that its ticks are homogeneous in time and they do not degenerate. The result is not significantly affected if we use a POVM with non-maximal localization, except that the  peaks in $P_c(t, \varphi)$ are less sharp. 

 For $\mu \neq 0$, the system still functions as a clock for times much smaller than $T_q$. The clock accuracy is $\tau = 2 \pi r^2 \omega_{\xi}/\xi$. As $t$ increases, the dispersion of the wave-packets makes the ticks non-distinguishable at the time-scale $T_q$.   The cumulative probability density $W(t)$ still increases broadly with $t$, but the accuracy is much worse than $\tau$. The wave-packet revival at time $t \sim T_{rec}$ does not increase clock accuracy, as the number of ticks  has become indefinite in the intermediate period. Hence, the time-scale $T_q$ sets an upper limit to the duration of the clock's functioning with maximum accuracy.

\section{Time-of-arrival probabilities in a rotating ring}\label{section5}
In this section, we generalize our previous analysis of time-of-arrival probabilities to a ring rotating with constant angular velocity $\Omega_D$.

The spacetime metric in the rotating frame reads
\bey
ds^2=-(1-\Omega_D^2r^2)dt^2 + dz^2 + d r^2 + 2\Omega_Dr^2d\varphi dt+r^2d\varphi^2,
\eey
in terms of the standard coordinates $(t, r, z, \theta)$ for cylindrically symmetric spacetimes. The Killing vector $\partial/\partial t$ is timelike only if $\Omega_D r < 1$  and we will restrict our analysis to this limit.

Following Ref. \cite{Letaw, Ambrus}, the quantization of a scalar field on the ring of constant $r$ and $z$ yields 
\begin{equation}
       \hat{\phi}(\varphi,t)=\sum_m\frac{1}{\sqrt{8\pi^2r|\omega_m|}}\left(\hat b_{m}e^{-i\tilde\omega_m t+im\varphi}+\hat b^\dagger_{m}e^{i\tilde\omega_m t-im\varphi}\right),
\end{equation}
where $m \in \Z$, and 
 $\tilde\omega_m=\pm\omega_m-m\Omega_D$ is the energy of the mode $m$ in the rotating system. Here, $\hat{b}_m$ and $\hat{b}^{\dagger}_m$ are annihilation and creation operators in the rotating reference frame. The rotating vacuum $|0\rangle_R$ is defined by $\hat{b}_m |0\rangle_R = 0$.

The field $\hat{\phi}$ is also expressed with the standard Minkowski modes, as in Eq. (\ref{minkphi}), with mode energy given by $\omega_m$. The key point is that the condition  $\Omega_D r < 1$ guarantees that $\tilde{\omega}_m \geq  0$ whenever $\omega_m \geq 0$ and conversely. This implies that the annihilation operators $\hat{a}_m$ associated with the Minkowski modes also annihilate the rotating vacuum. Hence, the rotating vacuum and the Minkowski vacuum coincide.

This is a profound difference from the Rindler vacuum relevant to linearly accelerated observers, which is physically inequivalent from the Minkowski vacuum. This inequivalence gives rise to the Unruh effect \cite{Unruh2, review}. While there is no inequivalent vacuum for rotating observers, many authors talk about a {\em rotational Unruh effect}, due to the non-trivial response of rotating detectors on the field vacuum \cite{Letaw81, Davies, Unruh, Levin, Moustos}.

In the present context, the rotational Unruh effect is not directly related the time-of-arrival problem in the sense that it does not affect the time-of-arrival probability distribution. It is manifested in a change of the background noise, the constant term $P_0$ in the detection probability that depends neither on the initial state nor on the location $\varphi$ of the detector and is constant with respect to time and corresponds to the false clicks of the detector. This term becomes $\Omega_D$-dependent: $ P_0(\Omega_D) = C \sum_{m}\tilde R(\omega_m - m \Omega_D,m)/(r\omega_m)$, where $C$ is a normalization constant. This sensitivity to the angular velocity of the detector is what leads to the correlation with the rotational Unruh effect.

 \color{black}

We can quantify the noise in terms of the ratio 
\bey
\eta = \frac{P_0(\Omega_D)}{P_0(0)}  = \frac{\sum_{m} \omega_m^{-1} \tilde R(\omega_m - m \Omega_D,m)}{\sum_{m} \omega_m^{-1} \tilde R(\omega_m ,m)}.
\eey 
The noise increases with $\Omega_D r$, diverging as $\Omega_D r \rightarrow 1$. 

\begin{figure}[H]
    \centering
    \includegraphics[scale=0.7]{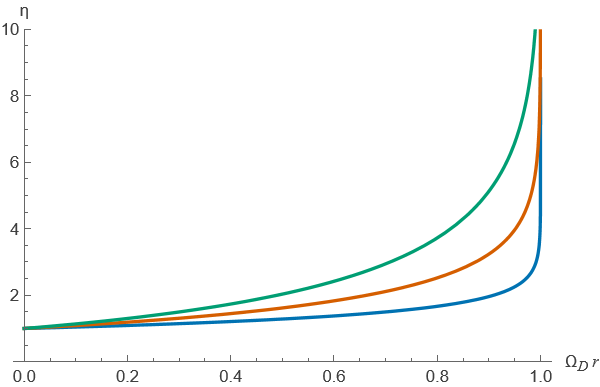}
    \caption[]{The normalized rotation induced noise $\eta = \frac{P_0(\Omega_D)}{P_0(0)}$ of Eq. (\ref{ffg}) as a function of $\Omega_Dr$ and for different values of the parameter $a$. }
    \label{fig:noise.png}
\end{figure}

We can obtain an explicit expression for a maximum localization detector and for $\mu = 0$. 
In this case,  $\tilde{R}(\omega_m, m) = e^{-a m} \theta(m)$ for some constant $a > 0 $ and we obtain
\bey
\eta = \frac{\log(1 - e^{-a (1 - \Omega_D r)})}{\log(1 - e^{-a })}. \label{ffg}
\eey 
The noise increases monotonically with $\Omega_D$---see Fig. \ref{fig:noise.png}. For $\Omega_D <<1$, $\eta -1$ increases linearly with $\Omega_D r$, while for $\Omega_D r \rightarrow 1$, $\eta$ diverges with $\log(1 - \Omega_Dr)$.



Having computed the noise term, it is straightforward to evaluate the probability distribution for the time of arrival from the remaining terms. The procedure is the same, the only difference being that the energy is given by $\tilde{\omega}_m$ rather than $\omega_m$. The probability density conditioned on detection is given by a direct analogue of Eq.(\ref{eq7}),

\bey
        P_c(t,\varphi) = \frac{B(\gamma)}{2 \pi r} \sum_{m,m'} \rho_{ps}(m,m') \sqrt{ \tilde v_m \tilde v_{m'}}\bra{m}\hat{L}_{\Omega_D}\ket{m'}e^{i(m-m')\varphi-i(\tilde\omega_m-\tilde\omega_{m'})t},
 \label{split}
\eey
where now the velocity of rotation $\tilde v_m = \frac{m}{\omega_mr}-r \Omega_D$ is evaluated on the rotating frame. The localization operator depends now on the angular velocity $\Omega_D$ and it is given by
  \bey
  \bra{m}\hat{L}_{\Omega_D}\ket{m'} = \frac{\tilde{R}\left(\frac{\tilde \omega_m+\tilde \omega_{m'}}{2},\frac{m+m'}{2}\right)}{\sqrt{\tilde{R}\left(\tilde \omega_m,m\right)\tilde{R}\left(\tilde \omega_m',m'\right)}}. \label{locop2}
  \eey
We observe that a maximum-localization detector remains so in presence of rotation.

Eq. (\ref{split}) has a natural normalization if the initial state has support only on the subspace where $\tilde{v}_m > 0$. Nonetheless, it can be extended to the full Hilbert space of the system. Assuming, for simplicity, a pure initial state $\psi_m$ and a maximum localization detector, we write
\bey
P_c(t,\varphi) = \frac{B(\gamma)}{2 \pi r} \left[ |{\cal D}_+(t, \varphi)|^2  + |{\cal D}_-(t, \varphi)|^2 +  {\cal D}_+(t, \varphi) {\cal D}^*_-(t, \varphi) + {\cal D}^*_+(t, \varphi){\cal D}_-(t, \varphi)\right],
\eey
where 
\bey
{\cal D}_{\pm}(t,\varphi) = \sum_{m} \Theta(\pm \tilde{v}_m) \psi_m \sqrt{|v_m|} e^{i m \varphi - i \tilde{\omega}_m t}.
\eey
Assume now that the initial state is symmetric in $m$, in the sense that $\psi_{-m} = \psi_m$, for $m > 0$.
If $\Omega_D r << v_m$, then ${\cal D}_{\pm}(t,0) = {\cal B}(t, \pm \Omega_D t)$ where ${\cal B}$ is the corresponding amplitude in absence of rotation. For an initial coherent state centered around $(\varphi, m) = (0, \xi)$ , ${\cal D}_{\pm}(t,0) \simeq {\cal B}(t, 0)e^{\pm i \xi \Omega_D t}$, and we find
\bey
P_c(t,0) = |{\cal B}(t, 0)|^2 \cos^2\left( \xi \Omega_D t\right).
\eey
There is an interference   phase $\xi \Omega_D t$ when the particles are measured at the point of entry $\varphi = 0$. This is one quantum analogue of the    \textit{Sagnac effect} of classical electromagnetism \cite{Sagnac}---see also Refs. \cite{Anandan, Hasselbach, Matos, Gautier}. For $t < T_q$,  ${\cal B}(t, 0)$ is peaked around times $t_n = 2 \pi n r/v_{\xi}$, where $n = 1, 2, \ldots$, so the interference phase increases as $2\pi n r^2 \omega_\xi \Omega_D$. The Sagnac phase is more clearly distinguishable at large $n$ provided the time remains smaller than $T_q$. 

There are  other manifestations of the Sagnac effect. For example, if the detector is located at a point $\varphi$, there is a phase difference of $\Delta \phi = 2 (\pi - \phi)$ between the detection time of the right- and left-moving modes. This difference is covered after $n$ rotations by $2\pi$, where
\bey
n = \frac{\pi - \varphi}{ \xi \Omega_D},
\eey
so at that time a stronger detection peak appears, corresponding to the coincidence in the detection of the two types of modes. Again, this effect can be observed  only if the required time $t$ is smaller than $T_q$.

\section{Multi-time measurements}\label{section6}

Our previous analysis straightforwardly applies to multi-time measurements. Consider, for example, the case of $n$ detectors, each located $\varphi = \varphi_i$, $i = 1,2, \ldots$. The joint probability density is
\bey
P_n(t_1, \varphi_1; t_2, \varphi_2; \ldots; t_n, \varphi_n) = \int ds_1 d \chi_1 \ldots ds_n d\chi_n     R_{(1)}(s_1, \chi_1) \ldots R_{(n)}(s_n, \chi_n)
\nonumber \\
\times G_{n,n}(t_1 - \frac{s_1}{2}, \vfi_1 - \frac{\chi_1}{2}  \ldots, t_n - \frac{s_n}{2}, \vfi_n - \frac{\chi_n}{2}  ; t_1 + \frac{s_1}{2}, \vfi_1 + \frac{\chi_1}{2} , \ldots, t_n + \frac{s_n}{2}, \vfi_n + \frac{\chi_n}{2} ), \label{probden4}
\eey
where $G_{n,n}$ is the correlation function with $n$ time-ordered and $n$ anti-time ordered entries
\bey
G_{n,n}(x_1, \ldots, x_n; x_1', \ldots, x_n') = \langle \psi| {\cal T}^*[\hat{\phi}(x'_1) \ldots \hat{\phi}(x'_n) ]
 {\cal T} [\hat{\phi}(x_n) \ldots \hat{\phi}(x_1)]|\psi\rangle. \label{correl}
\eey
In Eq. (\ref{correl}), we have assumed coupling through a composite operator $\hat{C}(x) = \hat{\phi}(x)$ as in the previous sections; ${\cal T}$ stands for time-ordering and ${\cal T}^*$ for anti-time-ordering.

For an initial state with a definite number of particles, all information about the initial state is encoded in the $n$-particle density matrix
\bey
\hat{\rho}_n(m_1, \ldots, m_; m'_1, \ldots m'_n) = \langle \psi|\hat{a}^{\dagger}_{m_1}\ldots \hat{a}^{\dagger}_{m_n} \hat{a}_{m_n} \ldots \hat{a}_{m_1}|\psi\rangle.
\eey

For $n = 2$, we find that the joint probability, normalized upon detection, is \cite{composite}
\bey
P_2(t_1, \varphi_1; t_2, \varphi_2) = \left(\frac{B(\gamma)}{2\pi r}\right)^2 Tr \left[\hat{\rho}_2 \hat{\Pi}^{(1)}_{t_1,\vfi_1} \otimes \hat{\Pi}^{(2)}_{t_2,\vfi_2}\right], \label{p2t}
\eey
where $\hat{\Pi}^{(i)}_{t,\vfi}$ are the positive operators (\ref{ptphi})---one for each detector. Eq. (\ref{p2t}) holds
modulo the noise term and a term that is strongly suppressed  for particle energies $E >> \delta_t^{-1}$, where $\delta_t$ is the characteristic scale for time-samplings, determined essentially from the detection kernel \cite{composite}. These conditions are generically satisfied for $\mu \neq 0$ and pose only mild restrictions for $\mu = 0$.

By Eq. (\ref{p2t}), the two detection events are dynamically independent: the measurement of one does not affect the other. They satisfy the Kolmogorov compatibility condition
\bey
\int dt_1 P_2(t_1, \varphi_1; t_2, \varphi_2)  = P_1(t_1, \vfi_1),
\eey
where $P_1(t, \vfi) = \frac{B(\gamma)}{2\pi r} Tr [\hat{\rho}_1 \hat{\Pi}(t, \vfi)]$ is the probability density (\ref{pcdd}). This property is specific to the coupling we have assumed, 
 by which particles are annihilated (absorbed) upon detection. The Kolmogorov condition does not apply when particles are detected through scattering. 

The probability density applies also to a different set-up that involves two particles, each on a different ring, and a simple detector on each ring. There the Hilbert space is ${\cal F} \otimes {\cal F}$. This set-up allows us to consider two distinct clocks, each localized in a different spacetime region, and analyze how quantum correlations induce correlations in clock readings. Note that if we take the probabilities (\ref{p2t}) to refer to different rings, there is no reason to assume particle identity, that is, that $\hat{\rho}_2$ is symmetric under particle exchange. Hence, the particle mass and the ring radius may be different in each component.

Of particular interest is the manifestation of quantum behavior in the joint probabilities (\ref{p2t}). Classical systems satisfy a locality criterion, known as {\em measurement independence} \cite{MI1, MI2, MI3}. This roughly asserts that the physical quantity that is being measured
in a detector is defined independently of any other quantity that is being measured in the
same set-up, and it is a prerequisite for the derivation of Bell's inequalities. 

In the present context, measurement independence implies the following two inequalities \cite{composite}
\bey
    P_1(t, \vfi)^2 &\leq& P_2(t,\vfi; t, \vfi), \label{jensen}\\
    P_2(t_1,\vfi_1; t_2, \vfi_2) &\leq& \sqrt{P_2(t_1,\vfi_1; t_1, \vfi_1)P_2(t_2,\vfi_2; t_2, \vfi_2)}. \label{CS}
\eey
 Consider a pure initial state $|\psi_1\rangle \otimes  |\psi_2\rangle$ and a detection kernel of maximum localization.
 Then,
 \bey
 P_1(t, \vfi) &=& \frac{B(\gamma)}{2\pi r} |{\cal A}_1(t, \vfi)|^2 \\
 P_2(t_1,\vfi_1; t_2, \vfi_2) &=& \left(\frac{B(\gamma)}{2\pi r }\right)^2\, |{\cal A}_1(t_1, \vfi_1)|^2  |{\cal A}_2(t_2, \vfi_2)|^2,
 \eey
 where ${\cal A}_i(t, \vfi) = \sum_m\psi_i(m) \sqrt{v_m}e^{im\varphi-i\omega_mt}$. We note that   inequalities (\ref{jensen}) and (\ref{CS}) are both saturated. The equations are still satisfied with mixed initial state or other localization operators, but less optimally. 
 
 Consider now an entangled initial state $\frac{1}{\sqrt{2(1+b)}} \left(|\psi_1\rangle \otimes  |\psi_2\rangle + |\psi_2\rangle \otimes  |\psi_1\rangle \right)$, where $b = |\langle \psi_1|\psi_2\rangle|^2$. Again, assuming maximum localization, we find
 \bey
  P_1(t, \vfi) &=&   \frac{B(\gamma)}{2\pi r } \left[|{\cal A}_1(t, \vfi)|^2+|{\cal A}_2(t, \vfi)|^2\right],\\
   P_2(t_1,\vfi_1; t_2, \vfi_2) &=& \left(\frac{B(\gamma)}{2\pi r}\right)^2 \frac{1}{2(1+b)} |{\cal A}_1(t_1, \vfi_1) {\cal A}_2(t_2, \vfi_2) + {\cal A}_1(t_2, \vfi_2) {\cal A}_2(t_1, \vfi_1)|^2
 \eey
 It is straightforward to show that the inequality  (\ref{jensen}) is violated when
 \bey
 |{\cal A}_2(t, \vfi)/{\cal A}_1(t, \vfi)|^2 \not\in [ 2\lambda  + 1 - 2\sqrt{\lambda(\lambda +1)},  2\lambda  + 1 +2\sqrt{\lambda(\lambda +1)}], \label{jensen2}
 \eey
 where $\lambda = \frac{2}{1+b} - 1 \in (0, 1]$.

Eq. (\ref{CS}) is violated if
\begin{equation}
    \frac{|{\cal A}_1(t_1, \vfi_1){\cal A}_2(t_2, \vfi_2)|}{|{\cal A}_1(t_2, \vfi_2){\cal A}_2(t_1, \vfi_1)|}\not\in[3-2\sqrt{2},3+2\sqrt{2}]. \label{CS2}
\end{equation}

We shall analyze some special cases. Taking $\mu = 0$ enables analytic calculations. For any initial state $|\psi\rangle = \sum_{m>0} \tilde{\psi}(m) |m\rangle$ that is strongly localized around zero, the associated amplitude ${\cal A}(t, \vfi)$ can be expressed through the Poisson method as
\bey
{\cal A}(t, \vfi) \simeq \sum_{n=-\infty}^{\infty} \psi(r(\vfi + 2 \pi n) - t),
\eey
where $\psi(x) = \int dk e^{ikx} \tilde{\psi}(k)$. So it suffices to determine an initial wave-function on $\R$ that is strongly localized at length scales much smaller than $r$. Then ${\cal A}(t, \vfi)$ is strongly peaked around $\bar{t}_n = r(\vfi + 2 \pi n)$ with a spread determined by the shape of $\psi(x)$.

First, we choose for $\psi_1(x)$ and $\psi_2(x)$ two Gaussian coherent states with different momenta $p_1$ and $p_2$: $\psi_i(x) = (2\pi \sigma^2)^{-1/4} e^{-\frac{x^2}{4\sigma^2}+i p_i x}$, for $i = 1, 2$. Then, in the vicinity of each peak of ${\cal A}(t, \vfi)$, 
\bey
\left|\frac{{\cal A}_1(t, \vfi)}{{\cal A}_2(t, \vfi)}\right| = 1,
\eey
and there is no violation either of Eq. (\ref{jensen}) or Eq. (\ref{CS}).

Consider now the case where $\psi_1(x) =  (2\pi \sigma^2)^{-1/4} e^{-\frac{x^2}{4\sigma^2}+i p x}$ and $\psi_2(x) = (x/\sigma) \psi_1(x)$ is orthogonal to $\psi_1$ ($b = 0$). Again, both amplitudes are peaked around $t_n = r(\vfi + 2 \pi n)$, and we find that
\bey\label{ineq}
|{\cal A}_2(\bar{t}_n + \tau, \vfi)/{\cal A}_1(\bar{t}_n + \tau, \vfi)|^2 = \frac{\tau^2}{\sigma^2},
\eey
\color{black}which violates the condition (\ref{jensen2}) for $|\tau|<\sqrt{3-2\sqrt{2}}\sigma$ or for $|\tau|>\sqrt{3+2\sqrt{2}}\sigma$.

\color{black}Taking $t_1 = \bar{t}_n + \tau_1$ in the vicinity of $\bar{t}_n$ and $t_2 = \bar{t}_m + \tau_2$ in the vicinity of $\bar{t}_m$, we find that 
\bey
  \frac{|{\cal A}_1(t_1, \vfi_1){\cal A}_2(t_2, \vfi_2)|}{|{\cal A}_1(t_2, \vfi_2){\cal A}_2(t_1, \vfi_1)|} = |\tau_2/\tau_1|.
\eey
Hence, Eq. (\ref{CS}) is violated for $|\tau_2| < (3-2\sqrt{2}) |\tau_1|$, or $|\tau_1| < (3-2\sqrt{2}) |\tau_2|$. 
\color{black}

\begin{figure}[ht]
\centering
  \begin{subfigure}{.7\textwidth} 
  \centering
    \caption[]{Violation of inequality (\ref{jensen2}).}
  \includegraphics[width=.7\linewidth]{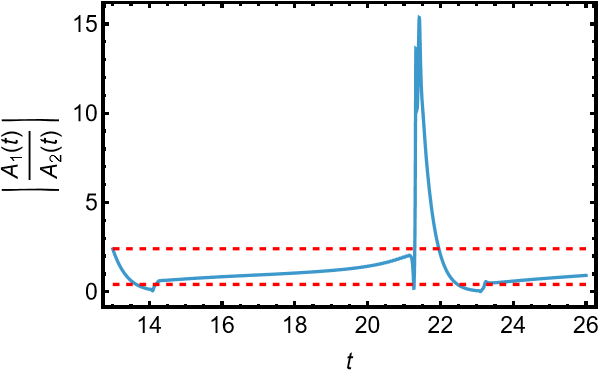}
    \end{subfigure}

  \begin{subfigure}{.5\textwidth} 
  \centering
  \caption[]{Violation of inequality (\ref{CS2}), lower bound.}
  \includegraphics[width=.8\linewidth]{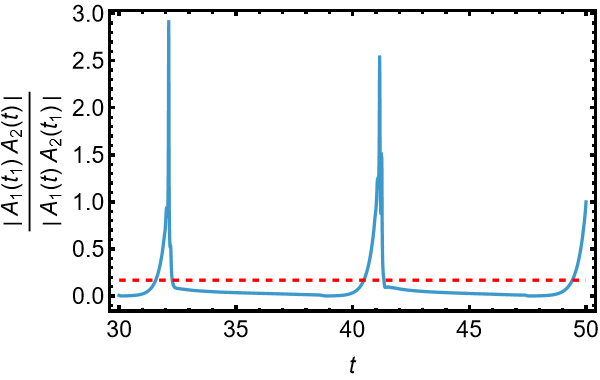}
  \end{subfigure}%
\begin{subfigure}{.5\textwidth}
\centering
    \caption[]{Violation of inequality (\ref{CS2}), upper bound.}
\includegraphics[width=.8\linewidth]{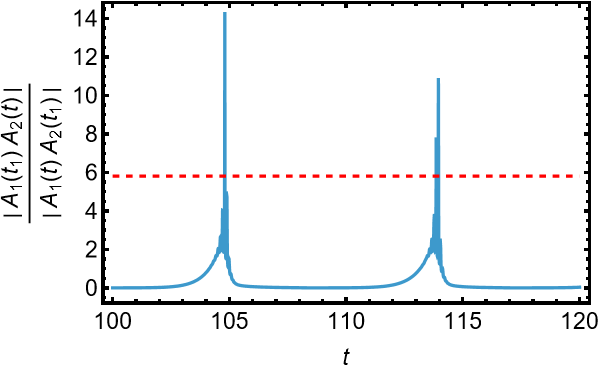}
  \end{subfigure}
    
\caption{Violation of measurement independence for massive particles. The initil state is  a superposition of coherent Gaussian states with different mean momenta $z_1=1005$, $z_2=995$, $M=1000$, $a=10$, $r=1$, $t_1=50$.}
\label{MIviolation1}
\end{figure}

For $\mu \neq 0$, we evaluate the amplitudes numerically. It turns out that---unlike the massless case---superpositions of Gaussian coherent states lead to violation of measurement independence---see Fig. \ref{MIviolation1}.

\section{Conclusions}\label{section7}
We have outlined our motivation and main results in the Introduction.
Here, we highlight their broader implications and potential applications.

A central result of this work is the construction of a relativistic quantum clock based on time-of-arrival measurements in a ring, derived  entirely within Quantum Field Theory. Because the broader framework is grounded on detector observables, it is directly sensitive to the local properties of spacetime. This makes it a natural tool for probing quantum spacetime effects in conceptually important settings, including regions near horizons such as those of black holes or Rindler observers.

In particular, the ring model also provides a controlled setting for studying the macroscopic records of  quantum clocks. Detection statistics depend on the quantum state of the field, allowing one to explore the roles of entanglement, coherence, and correlations in clock readings. At the same time, relativistic effects such as time dilation are incorporated at a fundamental level, enabling the study of their interplay with quantum measurement.

More generally, our results are a stepping stone towards a systematic investigation of measurable QFT phenomena in rotating frames. The ring geometry offers the simplest setting in which such effects can be analyzed. In particular, the appearance of rotation-induced noise—interpretable as a rotational Unruh effect—suggests that higher-order correlations and coherence properties of quantum fields may encode observable signatures of non-inertial motion.

These directions are part of the unified operational framework provided by the QTP approach, with applications to quantum clocks, relativistic quantum information, and quantum fields in gravitational and non-inertial environments.

\section*{Acknowledgements}

The authors were supported by the COST Action CA23115 "Relativistic Quantum Information".

\phantomsection

\end{document}